\documentclass[12pt]{article}
\usepackage[utf8]{inputenc}
\usepackage[english]{babel}
\usepackage{amssymb,graphicx}
\pdfoutput=1
\usepackage{graphics}
\usepackage{amsmath,amsfonts}
\usepackage{fullpage}
\usepackage{wrapfig} 
\usepackage{braket}
\usepackage[colorinlistoftodos]{todonotes}
\usepackage{multirow}

\newcommand{\bea}{\begin{eqnarray}}
\newcommand{\ena}{\end{eqnarray}}
\newcommand{\eq}{\begin{eqnarray}}
\newcommand{\en}{\end{eqnarray}}

\title{\vspace{-2cm} 
\begin{flushright}
{\normalsize INR-TH-2020-044}
\end{flushright}
\vspace{0.5cm} 
Constraints on CP-odd ALP couplings from EDM limits of fermions}
\author{D.~V.~Kirpichnikov$^1$\thanks{{\bf e-mail}: kirpich@ms2.inr.ac.ry}, Valery E.~Lyubovitskij$^{2,3,4,5}$ \thanks{{\bf e-mail}: valeri.lyubovitskij@uni-tuebingen.de }, Alexey S.~Zhevlakov$^{4,6,7}$\thanks{{\bf e-mail}: zhevlakov1@gmail.com }\vspace{.2cm}\\
\normalsize \it $^1$ Institute for Nuclear Research of the Russian Academy 
of Sciences, 117312 Moscow, Russia\\
\normalsize \it $^{2}$ 
\normalsize \it Institut f\"ur Theoretische Physik, Universit\"at T\"ubingen, \\ \normalsize \it Kepler Center for Astro and Particle Physics, \\
\normalsize \it Auf der Morgenstelle 14, D-72076 T\"ubingen, Germany\\
\normalsize\it $^2$ Institute for Nuclear Research of the Russian Academy
of Sciences, \\  
\normalsize \it $^{3}$ Departamento de F\'\i sica y Centro Cient\'\i fico 
Tecnol\'ogico de Valpara\'\i so-CCTVal, \\
\normalsize \it \hspace*{.5cm} Universidad T\'ecnica Federico Santa Mar\'\i a,
\normalsize \it Casilla 110-V, Valpara\'\i so, Chile \\
\normalsize \it $^{4}$  Department of Physics, Tomsk State University, 
634050 Tomsk, Russia \\
\normalsize \it $^{5}$ Tomsk Polytechnic University, 634050 Tomsk, Russia \\
\normalsize \it $^{6}$  Bogoliubov Laboratory of Theoretical Physics, \\
\normalsize \it Joint Institute for Nuclear Research, 141980 Dubna, Russia\\
\normalsize \it $^{7}$ Matrosov Institute for System Dynamics and \\
\normalsize \it Control Theory SB RAS Lermontov str., 134, 664033, Irkutsk, Russia
}
\date{}
\begin{document}
\maketitle

\begin{abstract}
We discuss constraints on soft CP-violating couplings of axion-like 
particles  with photon and fermions by using data on electric 
dipole moments of Standard Model particles.  
In particular, for the ALP leptophilic scenario
we derive bounds on CP-odd  ALP-photon-photon  coupling
from data of the {\tt ACME} Collaboration on electron EDM. 
We also discuss prospects of the Storage Ring experiment to constrain 
the ALP-photon-photon coupling from data on proton EDM
for the simplified  hadrophilic interactions of ALP.
The regarding 
constraints from experimental bounds on the muon and neutron EDMs are weak. 
We set constraint on the CP-odd ALP coupling with electron and derive 
bounds on combinations of coupling constants, which involve soft CP-violating terms. 
\end{abstract}

\section{Introduction}

Since resolving strong CP violation problem using Peccei-Quinn (PQ) 
mechanism~\cite{Peccei:1977hh} the axion-like particles (ALPs) 
proposed by Weinberg and Wilczek~\cite{Weinberg:1977ma,Wilczek:1977pj} 
play important role in hadron phenomenology and searching for 
New Physics (NP) beyond Standard Model 
(SM)~\cite{Alekhin:2015byh,Castillo-Felisola:2015ema}. 
In this vein the important step 
was formulation of the effective Lagrangian approach with explicit 
manifestation of the invisible axion~\cite{Georgi:1986df}. 
In particular, Lagrangian involving couplings of axion with SM 
gauge fields and fermions has been proposed. It was shown that 
the couplings of axion with SM gauge fields ($G=g,W,B$) 
are generated using anomalous coupling of the ALP to $G\tilde G$ 
gauge field currents, where $G$ and $\tilde G$ are generic strength 
of gauge field and its dual. 
In particular, the part describing the coupling of ALPs with 
photons and fermions $\psi=e,\mu,p$ 
reads~\cite{Georgi:1986df,Bauer:2017ris}     
\begin{eqnarray}
\mathcal{L} &\supset& \frac{1}{2} (\partial^\mu a)^2
- \frac{m_a^2}{2} \, a^2 
+ \frac{g_{a \gamma \gamma}}{4} \, a \, 
F_{\mu\nu} \tilde{F}^{\mu \nu} + 
\, \sum_{\psi=e, \mu, p} a g_{a \psi \psi} 
\, \bar{\psi} \, i\gamma_5 \, \psi \,, 
\label{LagrAggAff}
\end{eqnarray} 
where 
$g_{a \gamma \gamma} =  c_{a\gamma\gamma}/ \Lambda$
and 
$g_{a \psi \psi}=  c_{\psi\psi} \, m_f/\Lambda$
are the couplings of ALP with photons and fermions, 
$\Lambda$ is the NP scale, which is much larger 
than the electroweak scale $\Lambda_{\rm EW}$:  
$\Lambda \gg \Lambda_{\rm EW}$.  
One should stress that the coupling of axion with SM fermions 
are suppressed by $1/\Lambda$. Such coupling can be generated 
from the coupling of axion to the scalar fields (dimension-5 operator) 
after spontaneous breaking of electroweak symmetry~\cite{Bauer:2018uxu}.  We demonstrate
inducing of that coupling in Appendix.
  
In addition to this CP-even coupling, let us consider 
CP-odd coupling of ALP with photons  
\begin{eqnarray}
\mathcal{L}^{C\!P \!\!\!\!\!\!\backslash}_{a\gamma\gamma} \supset 
\frac{\bar{g}_{a \gamma \gamma}}{4} \, a \, F_{\mu\nu} \, F^{\mu\nu}\,,
\label{LagrAggAff_CP}
\end{eqnarray}
where $\bar{g}_{a\gamma\gamma}$ has dimension of GeV$^{-1}$. 
Such coupling was recently discussed in Ref.~\cite{Irastorza:2018dyq}. 
On the other hand, ALP is accompanied by a scalar field, dilaton $\phi$, 
in extra dimension theories. In particular, these degrees of 
freedom play important role in phenomenology of black holes and 
hadrons~\cite{Gibbons:1987ps,Garfinkle:1990qj,Shapere:1991ta,Gutsche:2011vb}. 
Coupling of the dilaton with photons has similar structure as the 
CP-odd one for the axion: 
$\mathcal{L} \supset \frac{g_{\phi \gamma \gamma}}{4} \,  
\phi \, F_{\mu\nu} F^{\mu \nu}$. 
Note, analogous couplings with two photons in case of light scalar mesons 
$f_0(600)$ and $a_0/f_0(980)$ have been studied in 
Refs.~\cite{Faessler:2003yf,Giacosa:2007bs,Branz:2008ha}.  
The coupling of the dilaton with fermions has Yukawa type, 
which is manifestly CP invariant: 
$\mathcal{L} \supset \bar{g}_{\phi \psi \psi} \, \phi \, \bar{\psi} \, \psi$. 
Note, the dilaton plays the role of the Nambu-Goldstone-like boson responsible 
for spontaneous breaking of conformal/scale invariance~\cite{Clark:1986gx}.
Its mass is expected below the typical conformal symmetry breaking scale 
$m_\phi \lesssim 1/g_{\phi \gamma\gamma}$. 

Constraints on $\phi\gamma\gamma$ coupling from collider experiments are 
widely discussed in the literature~ \cite{Ahmed:2019csf,LiuLiJia:2019kye,% 
Bandyopadhyay:2016fad,Megias:2015ory,Goncalves:2015oua,Efrati:2014aea,%
Jung:2014zga,Cox:2013rva,Barger:2011hu}  for the mass range 
$1\, \mbox{GeV} \lesssim m_\phi \lesssim 1$~TeV. In addition, authors of 
Ref.~\cite{Abu-Ajamieh:2017khi} provided a detailed analysis of light dilaton
scenarios ($1\, \mbox{keV} \lesssim m_\phi \lesssim 10$ GeV) and estimated  
the bounds on radion-photon-photon coupling $g_{\phi\gamma \gamma}$ from 
Supernova SN1987a, cosmology, Horizontal Branch stars, 
and beam-dump experiments. 
The latter analysis reveals an unconstrained window below 
$g_{\phi \gamma \gamma} \lesssim 10^{-5}$ GeV$^{-1}$ for the regarding 
mass range.
However, we note that emerging a CP violating coupling in the dilaton model 
$\mathcal{L} \supset g_{\phi \psi\psi} \, \phi \, \bar{\psi} 
\, i\gamma_5 \, \psi$ 
will require a proper recasting of the relevant bounds. 
That task however is beyond the scope of present paper. 
Instead,  we study CP-violating scenario~(\ref{LagrAggAff_CP}) 
for light sub-GeV pseudo-scalar particle and analyze in detail 
its implication for EDM physics of charged leptons and nucleons. 
In addition, for the certain ALP mass range we also set the limits 
on soft CP-violating coupling of ALP with electron: 
\eq 
\mathcal{L}^{C\!P \!\!\!\!\!\!\backslash}_{aee} \supset 
\bar{g}_{aee} a \bar{e} e\,.
\label{LagrAee_CP}
\en

In our previous paper~\cite{Kirpichnikov:2020tcf} we discussed 
NP phenomenology of hidden scalar, pseudoscalar, vector, and axial-vector 
particles coupled to nucleons and leptons, which could give contributions 
to different puzzles in particle physics (like proton charge radius, 
$(g-2)_\mu$, $^8$Be-$^4$He anomaly, electric dipole moments (EDMs) 
of SM particles). 

In the present paper we derive new limits on the couplings 
of ALPs with SM fermions using data on fermion EDMs. In particular, 
we consider the contribution of diagrams to fermion EDMs generated 
by the CP-even coupling of ALP with fermions and CP-odd coupling 
of ALP with photons. In this vein we do not require a universality 
of the coupling of ALP with leptons and quarks, which means that 
the limits on quark couplings with ALP are not necessarily applicable to
corresponding couplings in lepton sector. We note that constraints on combination of
CP-violating couplings from EDM physics are widely discussed in the literature. In particular, in Ref.~\cite{Dzuba:2018anu,Stadnik:2017hpa}   authors derived constraints on scalar and pseudoscalar
coupling constant combinations 
$|g_{aee} \bar{g}_{aee}|$ and $|g_{aee} \bar{g}_{app}|$
 from atomic and molecular EDM experiments for
the relatively wide range masses of ALP 
$10^{-6}~\mbox{eV} \lesssim m_a \lesssim 10^6$~eV.
 In Refs.~\cite{Yamanaka:2017mef,Yanase:2018qqq,Flambaum:2019ejc,Pospelov:2013sca} 
 authors discuss 
  constraints on CP-violating effective interactions $\bar{e}e \bar{N}N$
 from data on EDM of atoms and molecules.

The paper is structured as follows. 
In Sec.~\ref{SecYukawaBounds} we discuss the constraints on CP-even ALP-lepton 
couplings for the mass range of interest from $1$~MeV to $1$~GeV for leptophilic 
scenario of ALP interaction. 
 We also obtain the limits on ALP-photon-photon couplings
using data on electron and muon EDM. 
The expected bounds on ALP couplings from 
proton EDM are derived in Sec.~\ref{SectProtonMuonNeutron} for hadrophilic scenario of ALP interaction.  
 In Sec.~\ref{SecZfermOdd} we discuss 
bounds on CP-odd   couplings  associated with $a\gamma Z^0$ and $aee$ 
interaction. Combined bounds on products of ALP couplings are discussed in 
Sec.~\ref{SecCombBounds}.

\begin{figure}[t]
\begin{center}
\includegraphics[width=0.4\textwidth]{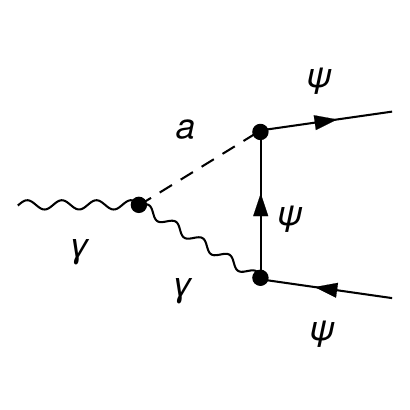}
\includegraphics[width=0.5\textwidth]{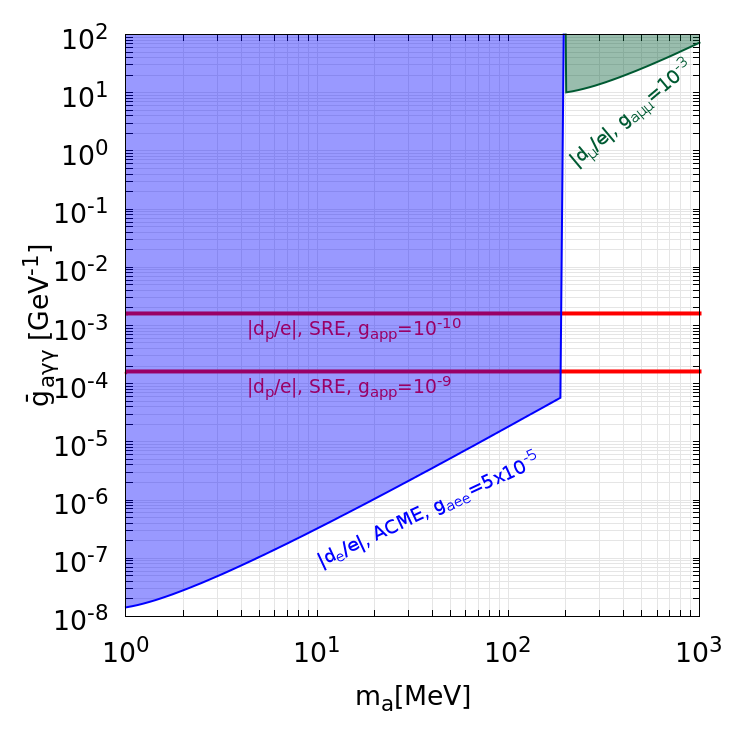}
%\vspace*{-.5cm}
\caption {Left panel: fermion EDM operator associated with CP-odd coupling
of  photon with ALP and CP-even interaction of ALP with SM fermions 
$\psi$, see, e.g., Eq.~(\ref{LagrAggAff}). Right panel:
Limits on $\bar{g}_{a \gamma\gamma}-m_a$ from various experiments. 
Blue region shows the parameter space of ALP at $90\%$ CL constrained by {\tt ACME}, 
that corresponds to the electron EDM limits 
$|d_e/e| < 1.1\times 10^{-29}$ cm and $g_{aee}\simeq 5.0\times 10^{-5}$ 
($c_{ee}/\Lambda \simeq 10^{-1}$ GeV$^{-1}$). Green region 
represents the current constraints on $\bar{g}_{a\gamma\gamma}$ 
from muon EDM, $|d_\mu/e| \lesssim 1.5 \times 10^{-19}$~cm, and benchmark coupling 
$g_{a\mu\mu} \simeq 10^{-3} (c_{\mu\mu}/\Lambda \simeq 10^{-2})$. 
Red solid 
lines are expected bounds on $\bar{g}_{a\gamma \gamma}$ for planned
sensitivity of {\tt SRE} to the proton EDM at the level of 
$|d_p/e| <  10^{-29}$ cm.  
\label{EDMloopDiagr}}
\end{center}
\end{figure} 

%\begin{figure}[t]
%\begin{center}
%\includegraphics[width=0.4\textwidth]{ALP_via_EDMe.png}
%\caption {Limits on $\bar{g}_{a \gamma\gamma}-m_a$ from various %experiments. 
%Blue line shows parameter space of ALP at $90\%$ CL constrained by {\tt %ACME} 
%that corresponds to the electron EDM limits 
%$|d_e/e| < 1.1\times 10^{-29}$ cm and $g_{aee}\simeq 5.0\times 10^{-5}$ 
%($c_{ee}/\Lambda \simeq 10^{-1}$ GeV$^{-1}$). Green solid line 
%represents the current constraints on $\bar{g}_{a\gamma\gamma}$ 
%from muon EDM, $d_\mu/e \lesssim 1.5 \times 10^{-19}$~cm, and benchmark %coupling 
%$g_{a\mu\mu} \simeq 10^{-3} (c_{\mu\mu}/\Lambda \simeq 10^{-2})$. 
%Red solid 
%lines are expected bounds on $\bar{g}_{a\gamma \gamma}$ for planing 
%sensitivity of {\tt SRE} to the proton EDM at the level of 
%$|d_p/e| <  10^{-29}$ cm.  
%\label{ACMElimits}}
%\end{center}
%\end{figure} 

\section{Constraints for leptophilic scenario 
\label{SecYukawaBounds}}
Let us consider Lagrangian describing the CP-odd coupling 
of ALP with SM photons and CP-even leptophilic interaction 
\begin{equation}
\mathcal{L} \supset  \frac{\bar{g}_{a \gamma \gamma}}{4}  
a F_{\mu\nu} F^{\mu \nu} + 
\sum_{l=e,\mu} i g_{a l l} \, a \, \bar{l} \gamma_5 l \,.
\label{LagrAggAee}
\end{equation}
These operators induce a finite contribution to the EDM of 
lepton. In the left panel of Fig.~\ref{EDMloopDiagr} we show  contribution
 of  1-loop diagram to the operator of fermion EDM. 
It must be point out that CP-even couplings $\mathcal{L}\supset 
\frac{g_{a\gamma \gamma}}{4} a F_{\mu\nu} \tilde{F}^{\mu\nu}$ 
don't generate lepton EDM operators at one-loop level. 
In Ref.~\cite{Kirpichnikov:2020tcf} authors 
showed that ALP Lagrangian~(\ref{LagrAggAee}) induces the lepton EDM, 
which has the following form
\begin{equation}
|d_l/e| = \frac{1}{16 \pi^2} \, \bar{g}_{a \gamma \gamma} 
\, g_{all} \, J\left(m_a/m_l\right),    
\label{GeneralEMDe}
\end{equation}
where function $J\left(m_a/m_l\right)$ for $m_a/m_l \gg 1$ 
can be approximated as
\begin{equation}
J\left(m_a/m_l\right) \simeq \frac{m_l^2}{3 m_a^2} 
\log\frac{m_a^2}{m_l^2},    
\end{equation}
and for light ALP, $m_a/m_l \ll 1$, it is given by
\begin{equation}
J\left(m_a/m_l\right) \simeq \frac{1}{2}\,. 
\label{JlightALP}
\end{equation}

Now, let us discuss existing constraints on CP-even ALP-fermion coupling  
$\mathcal{L}\supset i g_{all} a \bar{l}\gamma_5 l$ for the mass  
range of interest $1\, \mbox{MeV} \lesssim m_a \lesssim 1$~GeV.  
In particular, we refer to analysis of Ref.~\cite{Bauer:2017ris}  
on leptophilic coupling of ALP 
\begin{equation}
\mathcal{L} \supset \sum_{l=e,\mu,\tau} 
\frac{c_{ll}}{2\Lambda} \, (\partial_\mu a)\, 
\bar{l} \, \gamma^\mu \gamma^5 \, l
\label{cllALPee}
\end{equation} 
 It is appropriate to rewrite  coupling 
$g_{a ll}$ in Eq.~(\ref{LagrAggAee}) through
the  Yukawa-like term as follows $g_{a ll }=c_{ll} m_l/\Lambda$. 
Indeed, Eq.~(\ref{cllALPee}) on lepton mass shell implies that 
\begin{eqnarray}
\mathcal{L}\supset \sum_{l=e,\mu} \frac{c_{ll}}{\Lambda} \, 
m_l \, \bar{l} \, i\gamma_5 \, l \,.
\end{eqnarray} 
An author of Ref.~\cite{Bauer:2017ris} provided current limits 
on $c_{ll}/\Lambda$ from beam-dump experiments~\cite{Essig:2010gu} and  
{\tt BaBar} facility~\cite{TheBABAR:2016rlg} as well as from astro-particle
physics and cosmological  observations~\cite{Armengaud:2013rta}, assuming 
lepton universality of couplings, 
$c_{ee}\simeq c_{\mu\mu}\simeq c_{\tau\tau}$.      
In particular, in our estimate we use a benchmark conservative value 
$c_{ee}/\Lambda \simeq 10^{-1}$ GeV$^{-1}$ from coupling loop hole in 
the ALP mass range $1\, \mbox{MeV} \lesssim m_a \lesssim 200$ MeV. 
Finally, this implies $g_{a e e } \simeq 5 \times 10^{-5}$ for 
electron-ALP coupling. In addition, for the muon-ALP interaction we
take $g_{a\mu\mu}= 10^{-3}$ as a unconstrained 
benchmark coupling in the mass range 
$200\,~\mbox{MeV} \lesssim m_a \lesssim 1$~GeV, it corresponds to 
$c_{\mu\mu}/\Lambda = 10^{-2}\,\mbox{GeV}^{-1}$.

%\begin{figure}[t]
%\begin{center}
%\includegraphics[width=0.4\textwidth]{cllOverLambda_vs_ma.png}
%\caption { Exclusion limits for leptophilic ALP coupling 
%$c_{ll}/\Lambda\equiv g_{all}/m_l$ (see, e.g., Eq.~(\ref{cllALPee}) 
%for details) adapted from Ref.~\cite{Bauer:2017ris}
%\label{cllOverLambda}}
%\end{center}
%\end{figure} 

%\section{ALP bounds from lepton EDM
%\label{SecElectronEDM}}

We note that {\tt ACME} Collaboration~\cite{Andreev:2018ayy} 
set severe constraint on  electron EDM at $90\%$ CL, 
$|d_e/e| < 1.1 \times 10^{-29}\, \mbox{cm}$, or equivalently, 
\begin{equation}
  |d_e/e| \lesssim 5.5 \times 10^{-16}\, \mbox{GeV}^{-1}.
  \label{EDMeACME}
\end{equation}
Therefore, for $g_{a e e } \simeq 5 \times 10^{-5}$
it follows from Eqs.~(\ref{GeneralEMDe}), (\ref{JlightALP}),  
and~(\ref{EDMeACME}) that 
$
\bar{g}_{a \gamma \gamma} \lesssim 3.3 \times 10^{-9} \, \mbox{GeV}^{-1}
$
for $m_a \lesssim m_e$. Moreover for $m_a \gg m_e$ one has the following 
allowed limit on ALP-photon-photon coupling
\begin{equation}
\bar{g}_{a \gamma \gamma } \lesssim  5 \times 10^{-9}\, \mbox{GeV}^{-1}\times 
\frac{m_a^2}{m_e^2} \times \frac{1}{\log(m_a^2/m_e^2)}.   
\end{equation}
We note that existing limit on muon EDM, 
$|d_\mu /e| < 1.5 \times 10^{-19}$ cm, 
provides relatively weak bound 
\begin{equation}
\bar{g}_{a \gamma \gamma } \lesssim  3.6 \, \mbox{GeV}^{-1}\times 
\frac{m_a^2}{m_\mu^2} \times \frac{1}{\log(m_a^2/m_\mu^2)}.   
\end{equation}
for the benchmark coupling $g_{a\mu\mu} \simeq 10^{-3}$ and ALP 
masses in the range $200\,~\mbox{MeV}\lesssim m_a \lesssim 1$~GeV.
In the right panel of Fig.~\ref{EDMloopDiagr} we show   ALP parameter 
space constrained  by the experiments which are sensitive to  EDMs 
of leptons.

However, one remark should be added. 
For concreteness in our study we consider non-universal ALP coupling 
with leptons and quarks, $c_{ll} \neq c_{qq}$, which means that limits 
on $c_{qq}/\Lambda$ coming from meson 
decays~\cite{Essig:2010gu,Dolan:2014ska} are not directly applicable
to $c_{ll}/\Lambda$ bounds. 
We address hadrophilic constraints in Sec.~\ref{SectProtonMuonNeutron}.

\section{Constraints for hadrophilic scenario
\label{SectProtonMuonNeutron}}

\begin{figure}[t]
\begin{center}
\includegraphics[scale=0.5]{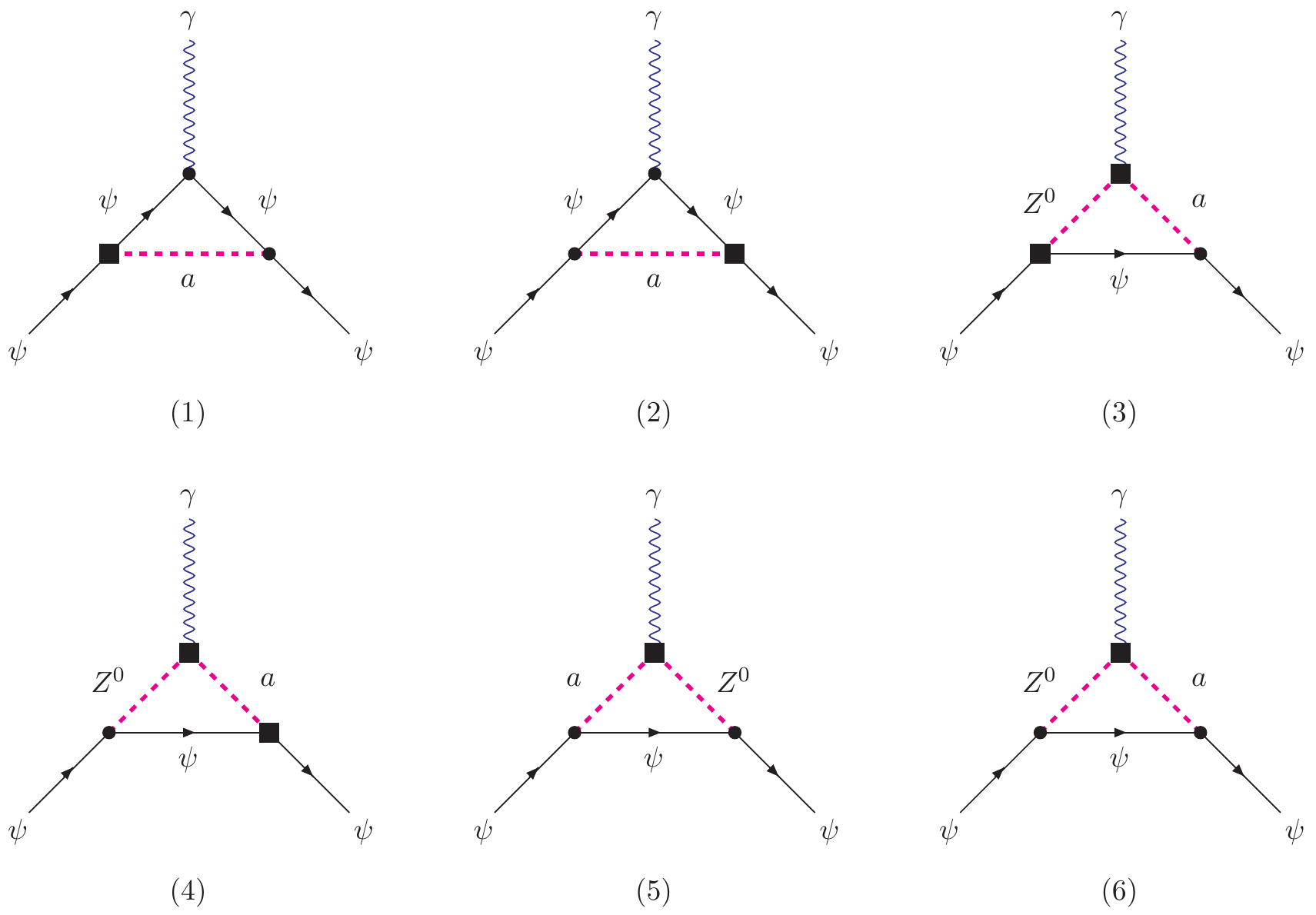}
\includegraphics[width=0.4\textwidth]{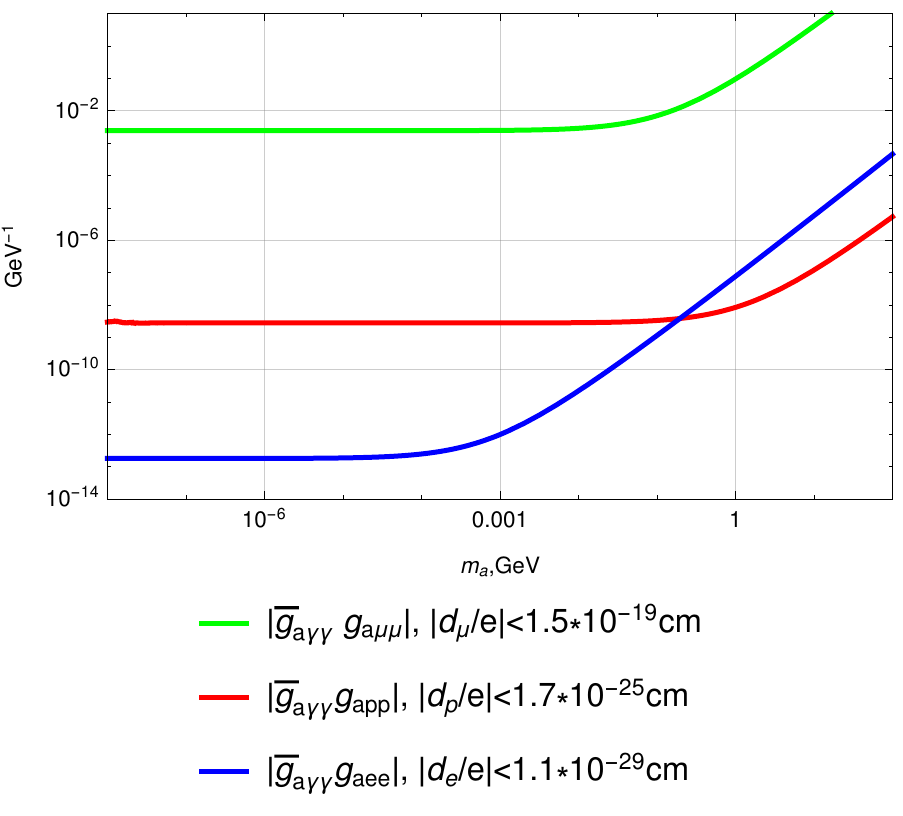}
\caption {Left panel: 
diagrams describing contribution of new particles to
fermion EDMs. Black boxes and dots represent CP-odd and CP-even vertices, 
respectively. In particular, in diagrams (1) and (2) black boxes   
correspond to $\bar{g}_{a\psi\psi} a \bar{\psi}\psi$ vertices,  
black rounds denote $i g_{a\psi\psi} a \bar{\psi}\gamma_5\psi$ vertices. 
In diagrams (3-6) black box (the $a\gamma Z $ vertex) corresponds 
to the interaction~(\ref{aFZ}).
Right panel: limits on $|\bar{g}_{a \gamma\gamma} g_{a\psi \psi}|-m_a$ ruled 
out at $90\%$ CL by experiments which are sensitive to measurements of EDM
of SM fermions. We address to Ref.~\cite{Bennett:2008dy} and  
Ref.~\cite{Graner:2016ses} for experimental constraints on muon and 
proton EDM respectively (for recent review 
see, e.g., Ref.~\cite{Kirch:2020lbo}).~\label{EDMCombCouplLimits}}
\end{center}
\end{figure} 
In this section we discuss the  constraints on the CP-even  
ALP-photon-photon couplings for the  following simplified
hadrophilic  scenario 
\begin{equation}
    \mathcal{L} \supset  \frac{\bar{g}_{a \gamma \gamma}}{4}  
a F_{\mu\nu} F^{\mu \nu} + 
\sum_{h=p,n} i g_{a hh } \, a \, \bar{h} \gamma_5 h \,.
\label{LagrAggAhh}
\end{equation}
 Let us consider first
 the prospects of the Storage Ring experiment ({\tt SRE}) 
(see, e.g., Ref.~\cite{Anastassopoulos:2015ura}) to probe ALP 
scenario with coupling to proton  
$\mathcal{L} \supset i g_{app} \, a \bar{p} \gamma_5 p$. In particular, 
in our analysis we consider the following typical bound on ALP-proton-proton 
coupling $g_{a pp} \lesssim 10^{-10}-10^{-9}$ for the mass range of interest 
$1\, \mbox{MeV} \lesssim m_a \lesssim 1$ GeV. 
In addition, the limit on $g_{app}$ is expected to be reasonable due to
ruled out limits on light pseudoscalar universal coupling with 
quarks~\cite{Dolan:2014ska} at the level of  $g_{aqq}\lesssim 10^{-8}$.  
The Storage Ring experiment is expected to be sensitive to the proton EDM
at the order of $|d_p/e| \lesssim 10^{-29}$ cm. This implies the 
following conservative bound on ALP-photon-photon interaction
\begin{equation}
\bar{g}_{a \gamma \gamma} \lesssim 1.6\times (10^{-4}-10^{-3})\, 
\mbox{GeV}^{-1}.
\end{equation}
The relevant expected limits for {\tt SRE} are shown in the right panel of
Fig.~\ref{EDMloopDiagr}. 
%\begin{figure}[tbh!]
%\begin{center}
%\vspace*{-.5cm}
%\hspace*{-3cm}
%\includegraphics[scale=0.5]{fig_axionEDM.pdf}
%\end{center}
%\vspace*{-5cm}
%\caption{Diagrams describing contribution of new particles to
%fermion EDMs. Black boxes and dots represent CP-odd and CP-even vertices,
%respectively. In particular, in diagrams (1) and (2) black boxes   
%correspond to $\bar{g}_{a\psi\psi} a \bar{\psi}\psi$ vertices,  
%black rounds denote $i g_{a\psi\psi} a \bar{\psi}\gamma_5\psi$ vertices. 
%In diagrams (3-6) black box (the $a\gamma Z $ vertex) corresponds 
%to the interaction~(\ref{aFZ}).}
%\label{fig:diag2}
%\end{figure} 
 The current limit on  neutron electric charge, 
$\bar{g}_{\gamma nn} < (-2\pm8)\times 10^{-22}e$ and its coupling
to ALP at the level of $g_{a n n}\lesssim 10^{-10}$ don't induce an 
experimentally favored constraints on $\bar{g}_{a \gamma \gamma}$ 
at one-loop level. 

\section{Constraints on ALP coupling with $Z^0$-boson and electron
\label{SecZfermOdd}}

It is worth mentioning, that one can also consider dimension-$5$ operator 
of ALP coupling with photon and $Z^0$-boson 
\begin{equation}
\mathcal{L} \supset \frac{\bar{g}_{a\gamma Z}}{2} a F_{\mu\nu} Z^{\mu\nu}, 
\label{aFZ}
\end{equation}
which provides a finite contribution to fermion EDM 
in association with CP-even,
$\mathcal{L}\supset  i g_{a\psi\psi} a \bar{\psi} \gamma_5\psi$, and CP-odd,
$\mathcal{L}\supset  \bar{g}_{a\psi\psi} a \bar{\psi} \psi$,  ALP interaction 
with fermions. In the left panel of Fig.~\ref{EDMCombCouplLimits} corresponding $1$-loop diagrams are 
labeled by (3), (4), (5), and (6). However, these diagrams generate 
sub-leading contributions to EDM of fermions due to suppression
factor $\sim 1/m_Z^2$, which is associated with $Z^0$-boson  internal line.
Therefore, CP-odd interaction~(\ref{aFZ}) doesn't induce viable constraint 
on $\bar{g}_{a\gamma Z}$ from EDM of fermions.

In our previous paper~\cite{Kirpichnikov:2020tcf} we derived
constraints on product of the couplings $\bar{g}_{aee}$ and $g_{aee}$  
from EDM bounds of electron. Corresponding $1$-loop diagrams, 
which induce electron EDM, are labeled by (1) and (2) in the left panel 
of Fig.~\ref{EDMCombCouplLimits}. However, it is instructive to obtain 
limits  on CP-odd coupling $\bar{g}_{aee}$    for certain  
values of benchmark coupling $g_{aee}$ and ALP mass $m_a$. 
Indeed,  one has the 
following estimate for the electron EDM~\cite{Bouchiat:1975qe}
\begin{equation}
\left|d_e/e\right| = \frac{\bar{g}_{aee} g_{a ee}}{8\pi^2 m_e} 
I(m_a/m_e),    
\end{equation}
where $I(m_a/m_e)$ can be approximated as 
\begin{equation}
I(m_a/m_e) = \left\{ 
\begin{array}{cc} 
\raisebox{0.5ex}
{$\frac{2m_\psi^2}{m_a^2} \log\left(\frac{m_a}{m_e}\right),$} & 
\raisebox{0.5ex}{$ m_a/m_e \gg 1,$} \\ 
\raisebox{-0.5ex}{$1,$} & 
\raisebox{-0.5ex}{$ m_a/m_e \ll 1.$} 
\end{array} \right.
\end{equation}
Therefore, for  $g_{aee} =5\times10^{-5}$ and the ALP mass range 
$1\, \mbox{MeV} \lesssim m_a \lesssim 200$~MeV  
one has the following conservative limits on~$\bar{g}_{aee}$ 
at 90$\%$ CL 
\begin{equation}
\bar{g}_{aee}   \lesssim 
\left\{ 
\begin{array}{cc} 
\raisebox{0.5ex}{$\!\!2.1\times 10^{-13}
\left(\frac{m_a}{m_e}\right)^2 \!\log^{-1}
\!\!\left(\frac{m_a}{m_e}\right) \!\!,$} & 
\raisebox{0.5ex}{$\!m_e \ll m_a ,$} \\ 
\raisebox{-0.5ex}{$4.3\times10^{-13},$} & 
\raisebox{-0.5ex}{$\!\!\!\!m_a \ll m_e.$}
\end{array} \right.
\end{equation} 
It is worth mentioning that the limit 
$\bar{g}_{aee} \lesssim 4.3 \cdot 10^{-13}$ for $m_a\gtrsim 20$~MeV is 
better than the bound on $\bar{g}_{aee}$ from non-resonant production 
of new scalars in Horizontal branch star core during helium burning
(for detail, see, e.~g. 
Ref.~\cite{Hardy:2016kme} and corresponding left panel in Fig.~5).

\section{Bounds on combination of couplings
\label{SecCombBounds}} 
In this section, for completeness, we summarize current 
reasonable constraints on combination of  couplings, which are 
ruled out by EDM of SM fermions. In the right panel of 
Fig.~\ref{EDMCombCouplLimits} 
we show regarding limits associated with muon, proton, and electron EDM. 
In particular, for relatively light ALP, $m_{a}\ll m_\psi$, 
one has the following scaling of the limit
\begin{equation}
\bar{g}_{a\gamma\gamma} g_{a\psi \psi} \lesssim 1.6 \times 10^{16}\, 
\mbox{GeV}^{-1} \times 
\left|\frac{d_\psi/e}{\mbox{cm}} \right|.
\label{lightMass}
\end{equation}
In addition for heavy ALP, $m_a \gg m_\psi$ one has 
\begin{equation}
\bar{g}_{a\gamma\gamma} g_{a\psi \psi} \lesssim 2.4 \cdot 10^{16}\, 
\mbox{GeV}^{-1} 
\frac{m_a^2}{m_\psi^2}\log^{-1}\left(\frac{m_a^2}{m_\psi^2}\right)     
\left|\frac{d_\psi/e}{\mbox{cm}} \right|
\label{heavyMass}
\end{equation}
One can see from Fig.~\ref{EDMCombCouplLimits} that most stringent 
constraint follows from electron EDM bound, 
$\bar{g}_{a\gamma\gamma} g_{a\psi \psi} \lesssim 10^{-13}\, 
\mbox{GeV}^{-1}$, as expected for $m_a \lesssim 1$ GeV. On the other hand,
the limit
on $\bar{g}_{a\gamma\gamma} g_{a\psi \psi}$ is ruled out
from proton EDM for $m_a \gtrsim 1$~GeV at the level of 
$\bar{g}_{a\gamma\gamma} g_{a\psi \psi} \lesssim 10^{-8}\, \mbox{GeV}^{-1}$. 
Finally, the combination of couplings associated with muon are feebly constrained 
due to the weak bounds on the muon EDM.

To conclude this section, we discuss known results on limits for the 
couplings from EDM bounds. In particular, we note that 
in Ref.~\cite{Dzuba:2018anu} authors 
derived similar constraints on scalar and pseudoscalar coupling constant combinations 
$|g_{e}^p g_{N}^s | \lesssim 10^{-16}$ from $^{199}\!$~Hg EDM experiments
for the typical masses $m_a \lesssim 10^6$~eV. These couplings are associated with the following Lagrangian 
$\mathcal{L} \supset  i  g_e^p a \bar{e} \gamma_5 e +  g_N^s a \bar{N} N$.
That combination connected with exchange
of an axion between the atomic electrons and the nucleus. 
Moreover, one can translate these couplings to our notations as follows, 
$g_e^p \equiv g_{aee}$ and $g_N^s \equiv \bar{g}_{app}$. 
In addition, authors of Ref.~\cite{Stadnik:2017hpa} provided constraints
on $|g_e^p  g_e^s| \lesssim 10^{-19}$ from atomic and molecular EDM experiment for $m_a \lesssim 10^6$~eV.  In our notation these couplings are 
$g_e^s \equiv \bar{g}_{aee}$ and $g_e^p \equiv g_{aee}$.
The relevant combination of coupling constant corresponds to  diagrams (1) 
and (2) in the left panel of Fig.\ref{EDMCombCouplLimits}. 
In Refs.~\cite{Yamanaka:2017mef,Yanase:2018qqq,%
Flambaum:2019ejc,Pospelov:2013sca} authors provided constraints 
on CP-violating contact interactions $\bar{e}e\bar{N}N$
from data on EDM of atoms and molecules.   

\section{Conclusion} 

In the present paper we derive constraints on soft CP-violating 
couplings of ALP from experimental data on  EDM bounds of SM fermions.  In
particular, we derive $90\%$ CL limit on CP-odd ALP coupling with photons
by taking into account EDM limits for leptons. 
That analysis is based on simplified phenomenological 
scenario of leptophilic ALPs in the mass range   $1\, \mbox{MeV} \lesssim m_a \lesssim 1$~GeV.  
We also obtain expected 
limits on CP-odd $a\gamma\gamma$  coupling for {\tt SRE} experiment on proton EDM which are associated with hadrophilic ALP interactions. 
We  calculate bounds on soft CP-violating ALP coupling with electron.

\section{Acknowledgements}
We would like to thank A.~Panin, S.~Gninenko, and 
N.~Krasnikov for fruitful discussions. 
The work of V.~E.~L. was funded
by ``Verbundprojekt 05P2018 - Ausbau von ALICE                                  
am LHC: Jets und partonische Struktur von Kernen''
(F\"orderkennzeichen: 05P18VTCA1),
``Verbundprojekt 05A2017-CRESST-XENON: Direkte Suche nach Dunkler 
Materie mit XENON1T/nT und CRESST-III. Teilprojekt~1 
(FZ.~05A17VTA)'', by ANID PIA/APOYO AFB180002 (Chile)
and by FONDECYT (Chile) under Grant No. 1191103, 
by the Tomsk State University Competitiveness Enhancement Program 
``Research of Modern Problems of Quantum Field Theory and Condensed Matter Physics''
and Tomsk Polytechnic University Competitiveness Enhancement Program (Russia). 
The work of A.~S.~Zh. was funded by the the Tomsk State University Competitiveness Enhancement 
Program ``Research of Modern Problems of Quantum Field Theory and Condensed 
Matter Physics'' (Russia).

\appendix 
\section{Generation of coupling of ALPs with SM fermions
\label{SecALPHiggsZferm}}

In this section we discuss a generation of the axion-$Z^0$ coupling. 
One could propose different scenarios of emergence of the axion-$Z^0$ 
coupling. E.g., it can be induced by the coupling of scalar field with 
axion (similar coupling of higher dimensions have been discussed in
Ref.~\cite{Bauer:2018uxu}). In particular, following ideas of
Ref.~\cite{Bauer:2018uxu}
one can consider the coupling of Higgs $\phi$- scalar fields with axion in the form induced
by dimension-5 operator
\begin{eqnarray}
{\cal L}_{a\phi} = \frac{G^{(5)}_{Zh}}{\Lambda}
\, \partial^\mu a\ (\phi^\dagger iD^\mu \phi)
\ \log(\phi^\dagger\phi/v^2)\,,
\end{eqnarray}
where $D^\mu$ is the covariant derivative including mixing of electroweak 
gauge fields and 
 $v = 246$ GeV is the Higgs vacuum condensate.
After realization of spontaneous breaking of electroweak symmetry and 
expressing combination of $B$ and $W^3$ fields
through $Z^0$ and $A$ one gets [using $G_{aZ^0} = G^{(5)}_{Zh} \log(1/2)]$:
\begin{eqnarray}
{\cal L}_{aZ^0} =
- \frac{e}{\sin(2\theta_W)} \, \frac{G_{aZ^0} v^2}{\Lambda} \, 
\partial^\mu a \, Z^0_\mu \,. 
\end{eqnarray}
By analogy with dark photon we can introduce the coupling of axion with 
SM fermions starting form mixing of axion and $Z^0$ boson parameterizing 
mixing parameter
$\bar\epsilon = \epsilon v^2/\Lambda$, where 
$\epsilon = e G_{aZ^0}/\sin(2\theta_W)$: 
\begin{eqnarray}
{\cal L}_{\rm mix} = - \epsilon \frac{v^2}{\Lambda} \partial^\mu a Z^0_\mu \,, 
\end{eqnarray}

Then we shift the $Z^0$ field as
\begin{eqnarray}\label{Z0_shift}
Z^0 \to Z^0 + \epsilon \frac{v^2}{\Lambda} \, \frac{\partial_\mu a}{M_Z^2} 
\,.
\end{eqnarray} 
It is clear that the kinetic term of the $Z^0$ field is unchanged 
after the shift~(\ref{Z0_shift}), 
while the kinetic term of the axion will get a very small correction
with can be eaten by the axion field redefinition:
\begin{eqnarray}
a \to a \cdot \Big[1 - \epsilon^2 v^4/(\Lambda^2M_Z^2)\Big]^{-1/2}
\end{eqnarray}
which results a negligible shift of the axion mass.
Like in case of dark photon,
the shift~(\ref{Z0_shift}) generates the couplings of the axion with 
SM fermions due to the shift in the coupling $Z^0\bar\psi\psi$: 
\eq 
{\cal L}_{Z^0\bar\psi\psi} \to {\cal L}_{Z^0\bar\psi\psi} + 
{\cal L}_{a\bar\psi\psi} \,,
\en 
where 
\eq 
{\cal L}_{Z^0\bar\psi\psi} = 
\frac{1}{2} \, Z^0_\mu \, 
\bar{\psi} \, [c_V^\psi \gamma^\mu
- c_A^\psi \gamma^\mu \gamma^5 ] \, \psi, \qquad 
{\cal L}_{a\bar\psi\psi} = 
\partial_\mu a \
\sum\limits_\psi \, \frac{g_\psi}{2 \Lambda} \,
\, \bar{\psi} \, \gamma^\mu \gamma^5 \psi \,,
\en 
%and 
%\eq 
%{\cal L}_{a\bar\psi\psi} = 
%\partial_\mu a \
%\sum\limits_\psi \, \frac{g_\psi}{2 \Lambda} \,
%\, \bar{\psi} \, \gamma^\mu \gamma^5 \psi \,. 
%\end{eqnarray} 
here $g_\psi$ is the coupling defined in consistency 
with original paper~\cite{Georgi:1986df} 
\eq 
g_\psi = - c_A^\psi \, \frac{\epsilon v^2}{M_Z^2} \,. 
\en 

%\clearpage 

%\vspace*{.25cm}
%\noindent Author Contributions: Conceptualization, D.V.K., V.E.L., A.S.Z.; 
                      Investigation, D.V.K., V.E.L., A.S.Z.; 
                      Writing—review and editing, D.V.K., V.E.L., A.S.Z.. 

%\conflictsofinterest{The authors declare no conflict of interest}

%\reftitle{References}

\end{document}